\documentclass[twocolumn,showpacs,preprintnumbers,amsmath,amssymb]{revtex4}
\usepackage{graphicx}% Include figure files
\usepackage{dcolumn}% Align table columns on decimal point
\usepackage{bm}% bold math

\begin{document}
\preprint{cond-mat/0601215}

\title{
Effect of the Vortices on the Nuclear Spin Relaxation Rate 
in the Unconventional Pairing States of the Organic Superconductor 
(TMTSF)$_2$PF$_6$
%Vortex excitation and pairing symmetry of quasi one dimensional organic superconductors via an NMR measurement
}% Force line breaks with \\

\author{M.~Takigawa$^{1}$}
\altaffiliation[current address: ]{Department of Physics, Okayama University, Okayama 700-8530, Japan}
\author{M.~Ichioka$^{2}$}
\author{K.~Kuroki$^{3}$}
\author{Y.~Asano$^{1}$}
\author{Y.~Tanaka$^{4}$}
% \altaffiliation[Also at ]{%
%Department of Applied Physics, Hokkaido University, Sapporo 060-8628, Japan
%}%Lines break automatically or can be forced with \\
%\author{Yasuhiro Asano}
\affiliation{%
$^{1}$Department of Applied Physics, Hokkaido University, Sapporo 060-8628, Japan \\
$^{2}$Department of Physics, Okayama University, Okayama 700-8530, Japan \\
$^{3}$Department of Applied Physics and Chemistry, %
the University of Electro-Communications, Chofu, Tokyo 182-8585, Japan \\
$^{4}$Department of Material Science and Technology, %
Nagoya University, Nagoya 464-8603, Japan}%

%\author{Masanori Ichioka}%
%\affiliation{%
%Department of Physics, Okayama University, Okayama 700-8530, Japan
%}%

%\author{Kazuhiko Kuroki}
%\affiliation{
%Department of Applied Physics and Chemistry, 
%the University of Electro-Communications, Chofu, Tokyo 182-8585, Japan
%}%

%\author{Yukio Tanaka}
%\affiliation{%
%Department of Material Science and Technology, 
%Nagoya University, Nagoya 464-8603, Japan
%}%

\date{\today}% It is always \today, today,

\begin{abstract}
This Letter theoretically discusses quasiparticle states and nuclear spin 
relaxation rates $T_1^{-1}$ in a quasi-one-dimensional 
superconductor (TMTSF)$_2$PF$_6$ 
under a magnetic field applied parallel to the conduction chains.
We study the effects of Josephson-type vortices on $T_1^{-1}$ 
by solving the Bogoliubov de Gennes equation for $p$-, $d$- or $f$-wave pairing 
interactions. 
In the presence of line nodes in pairing functions, 
$T_1^{-1}$ is proportional to $T$ in sufficiently low temperatures 
because quasiparticles induced by vortices at the Fermi energy relax spins.
We also try to identify the pairing symmetry of (TMTSF)$_2$PF$_6$.
\end{abstract}

% 74.70.Kn Organic superconductors  
% 74.20.Rp Pairing symmetries (other than s-wave) 
% 74.25.Op Mixed states, critical fields, and surface sheaths  
% 76.60.-k Nuclear magnetic resonance and relaxation  
% 76.60.Pc NMR imaging  

\maketitle

%\section{\label{sec:intro}Introduction}
Recently, much attention has been focused on superconductivity in 
a Bechgaard salt (TMTSF)$_2$PF$_6$~\cite{Bechgaard},  
where a superconducting phase appears next to a spin density wave phase 
in the pressure-temperature phase diagram~\cite{Jerome}.
So far a number of theories have proposed unconventional superconductivity 
in this quasi-one-dimensional (Q1D) organic superconductor: 
spin-singlet $d$-wave~\cite{Shimahara,Kino,Kuroki1999,DB}, 
triplet $p$-wave~\cite{Abrikosov,Hasegawa,Lebed} and 
triplet $f$-wave symmetries%
~\cite{Kuroki2001,Fuseya,Nickel,Tanaka,Kuroki2005,Fuseya2}.
Although pairing interaction along the chain direction is a common 
conclusion among these theories, 
the pairing symmetry itself has been controversial.
Experimentally, the spin-triplet pairing has been  
suggested from the unchanged Knight shift 
across the superconducting transition 
temperature $T_c$ in the NMR experiment~\cite{Lee} and the 
large enhancement of $H_{c2}$ exceeding the Pauli limit~\cite{Lee3}. 
The nuclear spin relaxation experiment is a powerful tool 
to identify the orbital part of the pairing function. 
The relaxation rate $T_1^{-1}$ should exhibit an exponential 
temperature($T$) dependence for fully gapped superconductors, while 
it is proportional to $T^3$ in the presence of line nodes in the gap. 
Such a theoretical analysis was applied to the TMTSF salts by 
Hasegawa and Fukuyama~\cite{HF}.%, and Takigawa {\it et al.}-->not needed ?
In (TMTSF)$_2$ClO$_4$, the relation $T_1^{-1}{\propto} T^3$ as well as 
the absence of the coherence peak were observed 
at zero magnetic field, which indicates 
the presence of line nodes~\cite{takigawaclo4}.
In (TMTSF)$_2$PF$_6$, however, Lee {\it et al.}~\cite{Lee} observed 
$T_1^{-1}\propto T$ in low temperatures under a magnetic field at 1.43 T.
Under a magnetic field, 
quasiparticles induced at the Fermi energy may cause the spin relaxation.
In fact, two of the present authors and other groups have shown that 
the relation $T_1^{-1}{\propto}T$ holds 
in a number of nodal superconductors~\cite{Takigawa1,Takigawa2,Silbernagel,Ishida}.  
In the NMR experiment by Lee {\it et al.}~\cite{Lee}, however, 
the magnetic field is applied parallel to the conduction chains.
At first sight, this experimental configuration seems 
to remove such quasiparticle contributions from $T_1^{-1}$
because the order parameter is mainly localized around the chain,  
and the vortices pass through the open space between the chains.
Thus order parameters along the chain are expected to be unchanged from 
their value at the zero magnetic field~\cite{Clem,Ichioka,Ichioka1995}. 
However, no theoretical study has ever tried to reveal effects
of such Josephson vortices on quasiparticle states and $T_1^{-1}$.

%%%%%%%%%%%%%%%%%%
\begin{figure}[t]
\includegraphics[width=6cm]{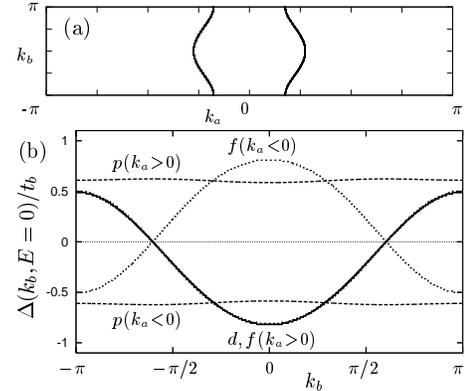}
\caption{\label{fig:k-E-D}
(a) Fermi surface in the Q1D tight-binding model ($E({\bf k}){=}0$) 
in the $k_a$-$k_b$ plane. 
$k_c$ dependence of the Fermi surface is negligible in this scale of the figure.
(b) The gap function $\Delta_0 \psi(k_a)$ along the Fermi surface 
for $p$-, $d$- and $f$-wave pairing symmetries at $T{=}0$ and $H{=}0$.
}
\end{figure}
%%%%%%%%%%%%%%%%%%

There are two main purposes to this Letter: (i) studying  
temperature dependences of $T_1^{-1}$ in the Q1D organic 
superconductor in the presence of vortices, 
and (ii) identifying the pairing symmetry of (TMTSF)$_2$PF$_6$.
For these purposes, we calculate quasiparticle states 
and order parameters self-consistently in the presence of the 
magnetic field along the chain 
direction based on the Bogoliubov-de Gennes (BdG) equation.
Three pairing symmetries
(i.e., $p$-, $d$- and $f$-wave) are assumed on a tight-binding lattice. 
We will show that the relation $T_1^{-1}{\propto} T$ holds 
at low temperatures for $d$- and $f$-wave symmetries.

We consider a three-dimensional tight-binding model, 
where the transfer integrals between the nearest neighbor 
sites are $-t_a$, $-t_b$ and $-t_c$ in the $a$, $b$ and $c$ directions, 
respectively. 
The band dispersion is given by 
$E({\bf k}){=}{-}2{t_a}\cos{k_a}{-}{2t_b}\cos{k_b}{-}{2t_c}\cos{k_c}{-}\mu$. 
To describe the electronic structure of 
(TMTSF)$_2$PF$_6$, we set $t_a:t_b:t_c{=}10:1:0.03$ 
and tune the chemical potential $\mu$ to keep the band filling at 
quarter filling (i.e., 
${\langle}n{\rangle}{=}0.5$).  
The Q1D Fermi surface for these parameters 
is shown in Fig.~\ref{fig:k-E-D}(a). 
We introduce a spin-singlet pairing 
interaction between the second nearest neighbor 
sites in the chain direction ($a$ axis) 
in the case of the $d$-wave symmetry~\cite{Tanuma}. 
For the $p$-wave ($f$-wave) pairing symmetry,  
 we introduce a spin-triplet pairing interaction~\cite{TakigawaP}
between the second (fourth) nearest sites in the $a$ direction.
The pairing functions result in $\psi(k_a){=}\cos{2k_a}$, $\sin{2k_a}$ and $\sin{4k_a}$ 
for the $d$-, $p$- and $f$-wave pairings, respectively. 
By applying the Fourier transformation in the $a$ axis, 
the Hamiltonian is given by 
\begin{eqnarray}
{\cal H}&{=}&\sum_{k_a,j,i,\sigma}
K_{k_a,i,j}a^\dagger_{j,\sigma} a_{i,\sigma} 
\nonumber \\ 
&+&\sum_{k_a,i}\{ 
 \Delta^\dagger_i \psi(k_a) a_{i,\downarrow} a_{i,\uparrow}
+\Delta_{i} \psi(k_a) a^\dagger_{i,\uparrow} a^\dagger_{i,\downarrow} \},
\label{eq:Ham}
\end{eqnarray}
where, $K_{k_a,i,j}{=}-\tilde{t}_{i,j} +\delta_{i,j}(-2t_a\cos{k_a}-\mu)$,  
 $a^{\dagger}_{i,\sigma}(a_{i,\sigma})$ 
is a creation (annihilation) operator of an electron with
spin $\sigma ({=}\uparrow$ or $\downarrow$), 
and $i{=}(i_b,i_c)$ is a site index in the $bc$ plane. 
For a magnetic field applied along the $a$ direction [${\bf H}{=}(H,0,0)$], 
the transfer integral in the $bc$ plane becomes
$\tilde{t}_{i,j}{=}t_{i,j}
{\exp}[{\rm i}\frac{\pi}{\phi_0}\int_{{\bf r}_i}^{{\bf r}_j}
{\bf A}({\bf r}) \cdot {\rm d}{\bf r} ]$, where $\phi_0$ is the flux quantum
and the vector potential is taken to be
${\bf A}({\bf r}){=}\frac{1}{2}{\bf H}{\times}{\bf r}$ 
in the symmetric gauge.

From Eq. (\ref{eq:Ham}), 
the BdG equation is derived for each $k_a$
\begin{eqnarray}
\sum_i
\left( \begin{array}{cc}
K_{k_a,i,j}         & D_{k_a,i,j} \\ 
D^\dagger_{k_a,i,j} & -K^\ast_{k_a,i,j}
\end{array} \right)
\left( \begin{array}{c} u_{\epsilon,j} \\ v_{\epsilon,j}
\end{array}\right)
&{=}&E_\epsilon
\left( \begin{array}{c} u_{\epsilon,i} \\ v_{\epsilon,i}
\end{array}\right),
\label{eq:BdG}
\end{eqnarray} 
where $u_{\epsilon,i}$ and $v_{\epsilon,i}$ are the wave functions belonging
to the eigenvalue $E_\epsilon$, and  
$D_{k_a,i,j}{=}{\psi}(k_a){\Delta}_i\delta_{i,j}$. 
The self-consistent equation for the pair potentials is given by 
\begin{eqnarray}
\Delta_i{=}U\sum_{\epsilon} u_{\epsilon,i}
v^\ast_{\epsilon,i} \psi(k_a)f(E_\epsilon) ,
\label{eq:delta}
\end{eqnarray}
where $f(E)$ is the Fermi distribution function  
and $U$ is the strength of the pairing interaction.  

We consider that two vortices accommodate in 
a unit cell of 20$\times$6 lattice sites in the $bc$ plane and 
form a triangular lattice. 
Note that the coherence length becomes anisotropic in the $bc$ plane,
(i.e., $\xi_b:\xi_c{\propto}t_b^{1/2}:t_c^{1/2}$).  
By introducing the quasimomentum of the magnetic Bloch state, 
the wave functions are calculated for sufficient quantities of
unit cells connected by the periodic boundary condition~\cite{Takigawa1}. 
We iterate the calculation of Eqs. (\ref{eq:BdG}) and (\ref{eq:delta}) 
alternately until a self-consistent solution is obtained. 
Using the self-consistent solution ($u_{\epsilon,i}$, 
$v_{\epsilon,i}$ and $E_\epsilon$), 
the local density of states (LDOS) is given by 
\begin{eqnarray} 
N(E,{\bf r}_i)
{=}\sum_{\epsilon}\{|u_{\epsilon,i}|^2 \delta(E-E_\epsilon)
+|v_{\epsilon,i}|^2 \delta(E+E_\epsilon) \}.
\end{eqnarray}
The screening current ${\bf J}({\bf r})$ is also calculated.
For instance, the current component in the $b$ direction is given by 
$J_{b}({\bf r}_i){\propto}{\rm Im} [ 
\tilde{t}_{i,i+\hat{b}}\sum_{\epsilon} \{ 
 u_{\epsilon,i} u^\ast_{\epsilon,i+\hat{b}}f(E_\epsilon)
+v^\ast_{\epsilon,i} v_{\epsilon,i+\hat{b}}(1-f(E_\epsilon)) \} ]$, 
where $i+\hat{b}$ denotes the nearest neighbor site to $i$ in
the $b$-direction~\cite{Takigawa1}. 
%%%%%%%%%%%%%%%%%
\begin{figure*}[bt]
\includegraphics[width=13cm]{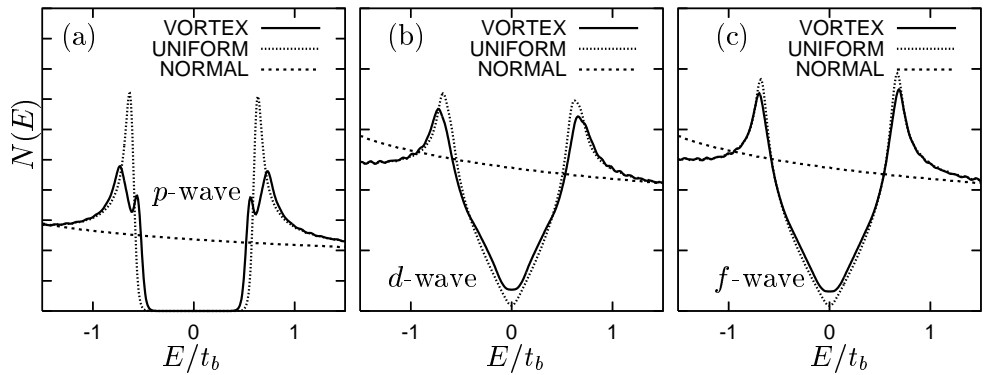}% Here is how to import EPS art
\caption{\label{fig:LDOS} 
The LDOS $N(E,{\bf r})$ at the site nearest to the vortex core 
is presented  at $T{=}0$ for the $p$-wave in (a), 
$d$-wave in (b) and $f$-wave in (c).
The LDOS in the vortex state is plotted with solid lines. 
The results in the uniform state at zero field and those in the normal state 
are shown with the dashed and dotted lines, respectively.
}
\end{figure*}
%%%%%%%%%%%%%%%%%
In the calculation, we take 
$U{=}{-}38t_b$ (giving $\Delta_0{=}2.4t_b$ and $T_c{=}1.05 t_b$) 
for $d$-wave pairing, 
$U{=}{-}15t_b$ ($\Delta_0{=}0.62t_b$ and $T_c{=}0.35 t_b$) 
for $p$-wave pairing, and  
%2006-05-11 $U{=}{-}30t_b$ ($\Delta_0{=}2.0t_b$ and $T_c{=}0.90 t_b$) 
$U{=}{-}27.5t_b$ ($\Delta_0{=}1.3t_b$ and $T_c{=}0.55 t_b$) 
for $f$-wave pairing with $\Delta_0{=}\Delta_i(H{=}0)$. 
Here $U$ is changed depending on pairing symmetries so that we have 
similar gap values in the three symmetries.
The ratio $\Delta_0/T_c$ depends on the anisotropy of $|\psi(k_a)|$. 
Figure~\ref{fig:k-E-D}(b) shows the superconducting gap 
$\Delta_0\psi(k_a)$ along the Fermi surface at $H{=}0$ for each symmetry.

First we briefly summarize the density of states (DOS) spectra $N(E,{\bf r}_i)$ 
in the absence of the magnetic field (i.e., in the uniform state) 
as shown by the dashed lines in Fig.~\ref{fig:LDOS}.
The dotted lines in Fig.~\ref{fig:LDOS} denote the DOS in the normal state.
In the $p$-wave symmetry, no nodes exist in the gap along the Fermi surface 
as shown in Fig.~\ref{fig:k-E-D}(b). 
Correspondingly, the full gap structure can be seen in the DOS as shown 
in Fig.~\ref{fig:LDOS}(a). 
Two peaks at $E{=}{\pm}0.65t_b$ reflect the large enhancement of the DOS 
at the gap edge.
In the $d$- and $f$-wave symmetries, the gap functions have 
line nodes at $k_a{=}\pm\pi/4$ on the Fermi surface. 
As a result, DOS shows a V-shaped gap structure at $H{=}0$, 
(i.e., $E$-linear DOS at low energies) 
as shown by the dashed lines in Figs.~\ref{fig:LDOS}(b) and ~\ref{fig:LDOS}(c).

%\section{\label{sec:DOS}Local Density of States around the vortex}

%%%%%%%%%
\begin{figure}[bt]
\includegraphics[width=6cm]{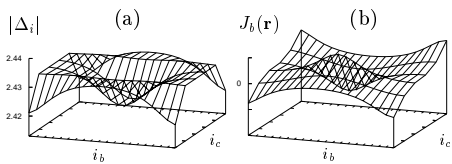}% Here is how to import EPS art
\caption{\label{fig:vortex} 
Spatial profile of the pair amplitudes and supercurrents in a unit cell 
are shown for the $d$-wave symmetry at $T{=}0$. 
In (a), the amplitudes of the pair potentials ($|\Delta_i|$) are presented.
In (b), the amplitudes of the $b$ component of 
the screening current ($J_b({\bf r})$) are presented.
}
\end{figure}
%%%%%%%%%%

Before turning to the DOS in the presence of vortices,
a typical spatial profile of the pair amplitude under the 
magnetic field should be discussed.
In Fig.~\ref{fig:vortex}(a), we show $|\Delta_i|$ 
for the $d$-wave symmetry for instance. 
Since the vortex center is located in the space between 
conduction chains, $|\Delta_i|$ is suppressed 
very slightly even near the vortex.
The screening current flows around the vortex as shown 
in Fig. \ref{fig:vortex}(b), 
where the $b$-component of the supercurrent is presented. 
Compared to the conventional vortex structure, 
the screening current flows in much wider region around the vortex. 
These features are typical characters of the Josephson vortex in 
layered superconductors~\cite{Clem,Ichioka,Ichioka1995}.

Second, we show the LDOS in the vortex state as presented 
by the solid lines in Fig.~\ref{fig:LDOS}, where 
we show $N(E,{\bf r})$ at the lattice site nearest to the vortex. 
In the fully-gapped $p$-wave symmetry, quasiparticles have bound states 
around the Josephson-type vortex with the energy 
near the full gap energy~\cite{Ichioka}. 
The two smaller peaks in LDOS around $E{\sim}{\pm}0.56t_b$ reflect
such bound states. 
The larger peaks around $E{\sim}{\pm}0.73t_b$ correspond to gap edges 
smeared by the magnetic field. 
In the $p$-wave, there is no quasiparticle states at the Fermi energy 
and the wide gap remains even in the presence of vortices.
Such behaviors are typically found in the quasiparticle states 
around the Josephson vortex in full gap superconductors.
In the $d$-wave [Fig.~\ref{fig:LDOS}(b)] and 
the $f$-wave [Fig.~\ref{fig:LDOS}(c)] symmetries, on the other hand, 
V-shaped superconducting gap is filled 
with low energy quasiparticles induced by vortices. 
Although suppression of $\Delta_i$ around the Josephson vortex 
is very small as shown in Fig.~\ref{fig:vortex}(a), 
vortices surely induce quasiparticles at the Fermi energy. 
Quasiparticles cannot form a zero-energy peak of the vortex core state
because of the small suppression of $\Delta_i$~\cite{Takigawa1,Takigawa2}. 
In both $d$- and $f$-wave, there is 
no bound states of quasiparticle in vortices because of line nodes. 
The calculated results in Fig.~\ref{fig:LDOS} show characteristic low energy 
quasiparticle excitations depending on the gap functions under the 
magnetic field.
%

%\section{\label{sec:NMR}Nuclear Magnetic Relaxation Rate}

Next we discuss the $T$-dependence of $T_1^{-1}$ on the 
basis of the calculated results 
in Fig.~\ref{fig:LDOS}.
The spin-spin correlation function 
$\chi_{\pm}({\bf r}_i,{\bf r}_{i'},{\rm i}\Omega_n)$ can be calculated
from the wave functions of the BdG equation. 
The nuclear spin relaxation rate is given 
by~\cite{Takigawa1,Takigawa2} 
\begin{eqnarray}
R({\bf r}_i,{\bf r}_{i'}) &{=}&
{\rm Im}\chi_{+,-}({\bf r}_i,{\bf r}_{i'},
{\rm i} \Omega_n \rightarrow \Omega + {\rm i}\eta)/(\Omega/T)|_{\Omega
\rightarrow 0}
\nonumber \\
&{=}&
 -\sum_{\epsilon,\epsilon'} 
[
 u_{\epsilon,i} u^\ast_{\epsilon',i'}
 v_{\epsilon,i} v^\ast_{\epsilon',i'}
%\nonumber \\ &&
 -v_{\epsilon,i} u^\ast_{\epsilon',i'}
  u_{\epsilon,i} v^\ast_{\epsilon',i'}
 ]
\nonumber \\ &&
\times \pi T f'(E_\epsilon) \delta(E_\epsilon - E_{\epsilon'}) . 
\label{eq:T1}
\end{eqnarray}
We choose ${\bf r}_i{=}{\bf r}_{i'}$ 
because the nuclear spin relaxation at a local lattice site is dominant. 
A relation $\delta(x){=}{\pi}^{-1}{\rm Im}(x-i\eta)$ is used 
to handle the discrete energy levels of the finite size calculation. 
We typically use $\eta{=}0.02t_b$. 
%%%%%%%%%%%%%%%
\begin{figure*}[bt]
\includegraphics[width=13cm]{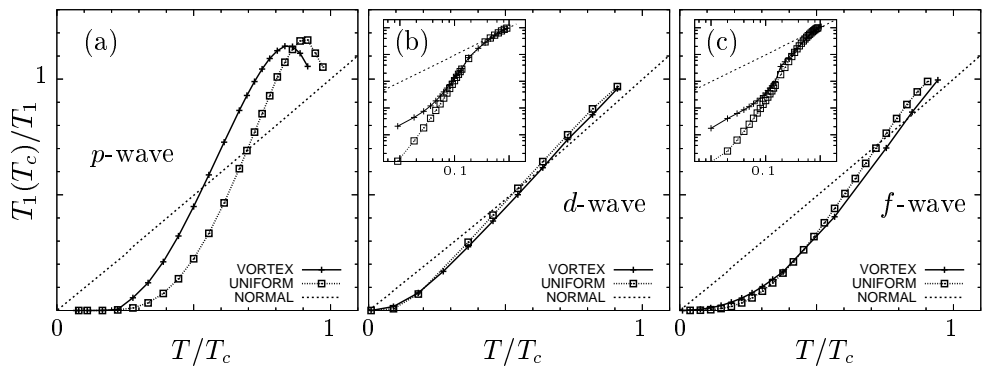}% Here is how to import EPS art
\caption{\label{fig:NMR} 
The nuclear spin relaxation rate $T_1^{-1}$ 
is shown as a function of temperature 
for (a) $p$-wave , (b) $d$-wave, and (c) $f$-wave symmetries.
Solid lines denote $T_1^{-1}$ at the lattice site nearest 
to the vortex core.
The results for the uniform state at $H{=}0$ and 
those in the normal state are plotted with 
dashed and dotted lines, respectively. 
Insets in (b) and (c) show the double logarithmic chart of the results.}
\end{figure*}
%%%%%%%%%%%%%%
%
In Fig.~\ref{fig:NMR},
we show $T_1^{-1}$ as a function of $T$ at the lattice site nearest 
to the vortex core with solid lines.
Although the relaxation time defined by 
$T_1({\bf r}){=}1/R({\bf r},{\bf r})$ depends on \textbf{r}, 
results in the following are almost independent of \textbf{r}.
In the normal state, we reproduce $T$-linear behavior of the Korringa law
as shown by the dotted lines in  Fig.~\ref{fig:NMR}. 
In the $p$-wave [Fig. \ref{fig:NMR}(a)],
$T_1^{-1}$  shows an exponential $T$ dependence for both the vortex and uniform states.
Effects of the vortices on $T_1^{-1}$ is almost negligible in low temperatures because 
there are no quasiparticle states around the Fermi energy even in the presence of
magnetic field as shown in Fig.~\ref{fig:LDOS}(a). 
By contrast, in the $d$-wave symmetry in Fig.~\ref{fig:NMR}(b) and 
the $f$-wave symmetry in Fig.~\ref{fig:NMR}(c), 
$T_1^{-1}$ deviates from $T^3$-behavior and becomes proportional to $T$ 
at low temperatures as observed in the 
experiment\cite{Lee} (see the double logarithmic chart in the insets). 
This implies that the 
quasiparticles at the zero-energy relax spins as they do in the normal state. 
Although the temperature at which $T_1^{-1}$ crosses over from 
$\propto T^3$ to $T$ seems to be somewhat below $T_c$ 
compared to the experimental result,\cite{Lee} 
this is mainly because we have taken $\Delta_0/t_b$ and $T_c/t_b$ to 
be larger than in the actual material due to the restriction in the 
numerical calculations. 
We have confirmed that this crossover temperature 
becomes higher for smaller $\Delta_0$, so that a more 
quantitative agreement with the experiment is expected for more realistic 
values of $\Delta_0/t_b$.

Finally we discuss the pairing symmetry in the actual TMTSF superconductor.
The magnetic field is applied along the $a$ axis in the experiment~\cite{Lee}
to remove the effect of the vortices on $T_1^{-1}$.
However our results show that the effect of the vortices 
is not negligible in unconventional superconductors with 
line nodes and that $T_1^{-1}$ is proportional to $T$ at sufficiently
low temperatures.
When large magnetic field is applied in other directions, 
the effect of vortices can be even more remarkable. 
In fact, $T_1^{-1}{\propto} T$ was also observed under a magnetic field
in the $b$ direction~\cite{Lee2}. 
In a very weak magnetic field, $T_1^{-1}$ is considered to be proportional 
to $T^{3}$ as observed in the experiment for 
(TMTSF)$_2$ClO$_4$~\cite{takigawaclo4}.
This further indicates that 
(i) the gap has line nodes that 
intersect the Fermi surface and 
(ii) the $T$-linear behavior at high magnetic field   
is caused by the effect of the vortices. 
We cannot 
discriminate between $d$- and $f$-waves solely from $T_1^{-1}$,
but if we combine this with the Knight shift~\cite{Lee} and the 
$H_{c2}$ measurement~\cite{Lee3}, triplet $f$-wave pairing seems to be 
the most promising candidate for the pairing state.

In summary, we have studied the quasiparticle states and 
the temperature dependence of the nuclear spin relaxation rate 
$T_1^{-1}$ in a quasi-one-dimensional organic superconductor (TMTSF)$_2$PF$_6$. 
We have considered the situation in which a magnetic field is 
applied along the chain direction as in the experiment.
As for the superconducting state, 
we have assumed three pairing symmetries ($p$-, $d$- and $f$-wave). 
Although the suppression of the order parameters 
near the Josephson vortex is very small, 
the Josephson vortex induces quasiparticles at 
the Fermi energy when the gap has line nodes 
($d$- and $f$-wave symmetries). 
As a result, the LDOS at the Fermi energy near the vortex 
becomes finite in the magnetic field.
Since $T_1^{-1}$ at low temperatures is very sensitive to 
the LDOS near the Fermi energy, 
$T_1^{-1}$ is proportional to $T$ in $d$- and $f$-wave symmetries.
On the other hand in the fully gapped case ($p$-wave symmetry), 
no quasiparticles is induced at the Fermi energy even in the magnetic 
field, so that the exponential dependence 
of $T_1^{-1}$ on $T$ remains unchanged.
We conclude that 
the relation $T_1^{-1} \propto T$ observed in (TMTSF)$_2$PF$_6$~\cite{Lee} 
is an evidence for line nodes in the gap.
Together with several experimental observation indicating spin-triplet pairing~\cite{Lee,Lee3}, 
our results suggest that the $f$-wave symmetry is the most promising candidate 
for Q1D organic superconductors.

%\begin{acknowledgments}
We thank K. Kanoda for useful discussion on the NMR experiment 
in organic superconductors. 
This work was supported by a Grant-in-Aid for the 21st Century 
COE ``Topology Science and Technology" in Hokkaido University.
K.K. acknowledges support from the Grant-in-Aid for 
Scientific Research from the Ministry of Education, 
Culture, Sports, Science and Technology of Japan.
%\end{acknowledgments}


\begin{thebibliography}{99}
\bibitem{Bechgaard}
%K. Bechgaard, {\it et al.},
K. Bechgaard, C.S. Jacobsen, K. Mortensen, H.J. Pedersen, and N. Thorup, 
Solid State Commun. {\bf 33} 1119 (1980).

\bibitem{Jerome}
%D. Jerome, {\it et al.},
D. Jerome, A. Mazaud, M. Ribault, and K. Bechgaard,
J. Phys. (Paris), Lett. {\bf 41}, L95 (1980). 

\bibitem{Shimahara}
H. Shimahara, 
J. Phys. Soc. Jpn. {\bf 58}, 1735 (1989).

\bibitem{Kuroki1999}
K. Kuroki and H. Aoki, 
Phys. Rev. B {\bf 60}, 3060 (1999).

\bibitem{Kino}
H. Kino and H. Kontani, 
J. Low. Temp. Phys. {\bf 117}, 317 (1999).

\bibitem{DB} 
R. Duprat and C. Bourbonnais, 
Eur. Phys. J. B {\bf 21}, 219 (2001).

\bibitem{Abrikosov}
A.A. Abrikosov, 
J. Low Temp. Phys. {\bf 53}, 359 (1983).

\bibitem{Hasegawa}
Y. Hasegawa and H. Fukuyama, 
J. Phs. Soc. Jpn. {\bf 56}, 877 (1987).

\bibitem{Lebed}
A.G. Lebed, 
Phys. Rev. B {\bf 59}, R721 (1999).

\bibitem{Kuroki2001}
%K. Kuroki, {\it et al.},
K. Kuroki, R. Arita, and H. Aoki, 
Phys. Rev. B {\bf 63}, 094509 (2001).

\bibitem{Fuseya}
%Y Fuseya, {\it et al.},
Y Fuseya, Y Onishi, H Kohno, and K Miyake, 
J. Phys. Cond. Matt. {\bf 14}, L655 (2002).

\bibitem{Nickel}
%J.C. Nickel, {\it et al.},
J.C. Nickel, R. Duprat, C. Bourbonnais, and N. Dupuis, 
Phys. Rev. Lett. {\bf 95}, 247001 (2005).

\bibitem{Tanaka}
Y. Tanaka and K. Kuroki, 
Phys. Rev. B {\bf 70}, 060502(R) (2004).

\bibitem{Kuroki2005}
K. Kuroki and Y. Tanaka, 
J. Phys. Soc. Jpn. {\bf 74}, 1694 (2005).

\bibitem{Fuseya2}
Y. Fuseya and Y. Suzumura, 
J. Phys. Soc. Jpn. {\bf 74}, 1263 (2005).

\bibitem{Lee}
%I.J. Lee, {\it et al.},
I.J. Lee, S.E. Brown, W.G. Clark, M.J. Strouse, 
M.J. Naughton, W. Kang, and P.M. Chaikin, 
Phys. Rev. Lett. {\bf 88}, 017004 (2002).

\bibitem{Lee3}
%I.J. Lee, {\it et al.},
I.J. Lee, M.J. Naughton, G.M. Danner, and P.M. Chaikin, 
Phys. Rev. Lett.  {\bf 78} 3555 (1997).

\bibitem{HF} 
Y. Hasegawa and H. Fukuyama, 
J. Phys. Soc. Jpn. {\bf 56}, 877 (1987).

\bibitem{takigawaclo4}
%M. Takigawa, {\it et al.},
M. Takigawa, H. Yasuoka, and G. Saito, 
J. Phys. Soc. Jpn. {\bf 56}, 873 (1987).
%M. Takigawa, H. Yasuoka, G. Saito, Y. Maniwa, and T. Takahashi, 
%Nuclear magnetic relaxation in 
%the organic superconductior (TMTSF)$_2$ClO$_4$, 
%in Nvel Superconductivity, ed by S.A. Wolf and V.Z. Kresin 141 (Plenum, New York, 1987), pp 141-9.

\bibitem{Takigawa1}
%M. Takigawa, {\it et al.},
M. Takigawa, M. Ichioka, and K. Machida, 
Phys. Rev. Lett. {\bf 83} 3057 (1999);
%M. Takigawa, {\it et al.},
M. Takigawa, M. Ichioka, and K. Machida, 
J. Phys. Soc. Jpn. {\bf 69}, 3943 (2000).

\bibitem{Takigawa2}
%M. Takigawa, {\it et al.},
M. Takigawa, M. Ichioka, and K. Machida, 
Phys. Rev. Lett. {\bf 90}, 047001 (2003);
%M. Takigawa, {\it et al.},
M. Takigawa, M. Ichioka, and K. Machida, 
J. Phys. Soc. Jpn. {\bf 73}, 450 (2004).

\bibitem{Silbernagel}
%B.G. Silbernagel, {\it et al.},
B.G. Silbernagel, M. Weger, and J.E. Wernick, 
Phys. Rev. Lett. {\bf 17}, 384 (1966).

\bibitem{Ishida}
%K. Ishida, {\it et al.},
K. Ishida, Y. Kitaoka, and K. Asayama, 
Solid State Commun. {\bf 90}, 563 (1994). 
%K. Ishida, {\it et al.},
K. Ishida, Y. Kitaoka, K. Asayama,
K. Kadowaki, and T. Mochiku, 
J. Phys. Soc. Jpn. {\bf 63}, 1104 (1994).

\bibitem{Clem}
J.R. Clem and M.W. Coffey, 
Phys. Rev. B {\bf 42}, 6209 (1990).  

\bibitem{Ichioka} 
M. Ichioka and T. Tsuneto, 
J. Low Temp. Phys. {\bf 96}, 213 (1994). 

\bibitem{Ichioka1995}
M. Ichioka, 
Phys. Rev. B {\bf 51}, 9423 (1995).  

\bibitem{Tanuma}
%Y. Tanuma, {\it et al.},
Y. Tanuma, K. Kuroki, Y. Tanaka, and S. Kashiwaya
Phys. Rev. B {\bf 64}, 214510 (2001).

\bibitem{TakigawaP}
%M. Takigawa, {\it et al.},
M. Takigawa, M. Ichioka, K. Machida, and M. Sigrist, 
Phys. Rev. B {\bf 65}, 014508 (2002).

%\bibitem{Tanuma}
%Y. Tanuma, K. Kuroki,and Y. Tanaka, unpublished.

\bibitem{Lee2}
%I.J. Lee, {\it et al.},
I.J. Lee, D.S. Chow, W.G. Clark, M.J. Strouse, M.J. Naughton, 
P.M. Chaikin, and S.E. Brown, 
Phys. Rev. B {\bf 68}, 092510 (2003).  


%\bibitem{Kanoda}
%K. Kanoda, private communications. 



\end{thebibliography}
\end{document}